\documentclass[]{spie}  

\usepackage{amsmath,amsfonts,amssymb}
\usepackage{graphicx}
\usepackage[colorlinks=true, allcolors=blue]{hyperref}

\DeclareMathOperator*{\argmin}{arg\,min}

\title{Using the generalised-optical differentiation wavefront sensor for laser guide star wavefront sensing}

\author[a,*]{Sebastiaan Y. Haffert}
\author[b]{Christoph U. Keller}
\author[c]{Richard Dekany}
\affil[a]{University of Arizona, 933 N. Cherry Ave., Tucson, AZ USA}
\affil[b]{Leiden University, Address, Leide, The Netherlands}
\affil[c]{California Institute of Technology, 1201 E. California Blvd, Pasadena, CA USA}

\authorinfo{Further author information: (Send correspondence to S.Y. Haffert)\\S.Y. Haffert: E-mail: shaffert@arizona.edu \\
{\noindent \footnotesize *NASA Hubble Fellow}
}

\begin{document} 
\maketitle

\begin{abstract}
Laser guide stars (LGS) are used in many adaptive optics systems to extend sky coverage. The most common wavefront sensor used in combination with a LGS is a Shack-Hartmann wavefront sensor (SHWFS). The Shack-Hartmann has a major disadvantage for extended source wavefront sensing because it directly samples the image. In this proceeding we propose to use the generalized-Optical Differentation Wavefront Sensor (g-ODWFS) a wavefront sensor for wavefront sensing of LGS. The g-ODWFS uses only 4 pixels per sub-aperture, has little to no aliasing noise and therefore no spurious low-order errors and has no need for centroid gain calibrations. In this proceeding we show the results of simulations that compare the g-ODWFS with the SHWFS.
\end{abstract}


\section{INTRODUCTION}
\label{sec:intro}  
One of the major limitations for ground-based telescopes is that these telescopes image through the Earth’s atmosphere. Due to turbulence in the atmosphere the light entering our telescope is aberrated which severely degrades the spatial resolution of the telescope, which in turn reduces sensitivity and astrometric precision. To combat the Earth’s atmosphere, we use adaptive optics (AO) where we actively control a deformable mirror to compensate for the effects of the turbulence. However, before we can correct the atmospheric turbulence we must first measure it. This is commonly done by a wavefront sensor. Because the atmosphere changes rapidly, we need to measure and correct at or above 1 kHz sample rate, limiting the technique to bright stars because those provide enough light at such speeds. To enable atmospheric correction on fainter stars, a bright artificial star can be created with a high-power laser \cite{foy1985feasibility}. Such a Laser Guide Star (LGS) uses a laser beam that back-scatters on certain atomic or molecular species high in the atmosphere which is then observed by the telescope\cite{1994JOSAA..11..263H}. In this work we consider sodium-based lasers that are in use at the Very Large Telescope (VLT) \cite{calia2010laser}. However, the results in this work are equally valid for lower-altitude Rayleigh LGS's.

Adaptive optics systems with an LGS have larger sky coverage than that available with natural guide star wavefront sensing \cite{le1998laser}. However, an LGS brings its own set of problems \cite{foy1985feasibility, Dam:06, Clare:07}. The sodium-based (Na) LGS take advantage of the high density sodium layer that is positioned roughly 90 km above the telescope. The sodium layer has a vertical extent of about 20 km. This thick layer creates an extended source if viewed off-axis. Currently, most LGS AO systems use a Shack-Hartmann wavefront sensor (SHWFS), which uses a micro-lens array in the pupil to segment the incoming wavefront into many sub-apertures. The slope of the wavefront in each sub-aperture can then be estimated from sub-aperture centroid motions and from integration of these slopes the full incoming wavefront can be reconstructed.

A problem for the SHWFS is that each microlens sees the LGS at a different off-axis angle, which creates an elongation of the LGS that varies across the pupil \cite{Dam:06}. The spot elongation depends on both the distance to the LGS and on how well the spot is resolved. This means that larger-aperture telescopes are hit twice, once because the distance from the LGS to the edge apertures is larger, but also because the diffraction-limit may be smaller and the LGS becomes better resolved, depending on other design choices. To keep the LGS well-sampled, a large number of pixels per measurement is necessary for the SHWFS. However, current technology limits the number of available pixels and therefore there is a trade-off between the pixel scale, the number of sub-apertures and the field-of-view. Increasing the pixel scale or truncation of the field-of-view leads to increased reconstruction errors \cite{Dam:06, thomas2006comparison}. The current design of the LGS system for the E-ELT uses an array of 80 by 80 sub-apertures each sampled by 10x10 pixels, which amounts to a total chip size of 800x800 pixels \cite{downing2014lgsd}. Such a detector also need to read out at sufficiently high speed so that the atmosphere has little time to evolve. This is technologically a very challenging detector and ESO has spent several years to develop such a detector \cite{downing2014lgsd,downing2018update}. The optimal detector size is actually larger with a 1600x1600 size, and is also under development \cite{gach2019c}. However, all these novel large chip detectors have lower read out speeds and more detector noise than current state-of-the-art EMCCD sensors with smaller sensor sizes \cite{agapito2018emccd}.



\begin{figure*}
	\begin{center}
        \includegraphics[width = \textwidth]{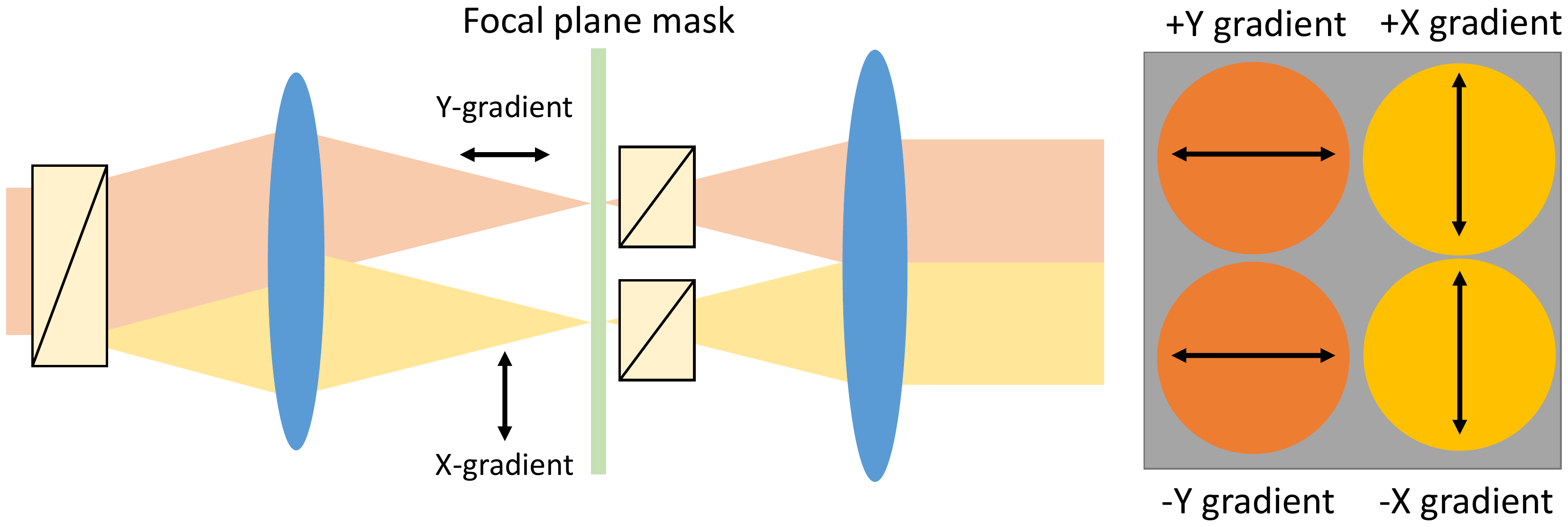}
	\end{center}
	\caption{A schematic of the g-ODWFS. The collimated beam is split into two by the first Wollaston prism, which are then focused onto the focal plane mask. One beam has a filter for the x-gradient, while the other beam has a filter for the y-gradient. After the focal plane mask both beams are split again, but in an orthogonal direction to the first beam splitter, and collimated by the second lens. The final image has 4 output pupils that encode the wavefront information.}
	\label{fig:odwfs_schematic}
\end{figure*}

Here we propose to use the generalised Optical Differentiation wavefront sensor (ODWFS) \cite{bortz1984wave, horwitz1994new}, which can circumvent many pixel sampling issues. The ODWFS is similar to Fourier-based sensors such as pyramid WFS but uses two modulating filters in the focal plane instead of a single mask. Each filter encodes either the x-gradient or the y-gradient of the wavefront aberrations. A schematic of this setup can be seen in Figure \ref{fig:odwfs_schematic}. The proposed ODWFS use amplitude filters, which lose light by definition. In the past few years the generalised-Optical Differentiation Wavefront Sensor (g-ODWFS) was create. The g-ODWFS is a version of the ODWFS based on polarization optics which is 100\% photon efficient \cite{haffert2016generalised, haffert2018sky}. A Wollaston prism splits the collimated input beam into two polarized beams, which are focused onto the different focal plane filters. For the g-ODWFS the focal plane filters are half-wave plates with a spatially varying fast-axis orientation, therefore the angle of polarization of the input beams will rotate by an amount that depends on its spatial position in the focal plane. A second Wollaston prism that is rotated by 45 degrees with respect to the first Wollaston prism is used as a polarization analyzer that creates four output beams, two for each gradient. The intensity difference between the two output beams per gradient depends on the angle of polarization, and therefore on the orientation of the fast axis of the focal plane mask where the beam hit it. We can measure for each position in the pupil what the angle of polarization is and derive where the ray hit the focal plane mask. From this we can measure the local tilt of the wavefront and reconstruct the wavefront aberrations. The g-ODWFS has been demonstrated in the lab and recently at the telescope as a wavefront sensor for point sources that can be used for astronomical adaptive optics \cite{haffert2018sky}.

\section{The Optical Differentiation Wavefront Sensor for LGS}
\label{sec:background}  

\begin{figure*}
	\begin{center}
        \includegraphics{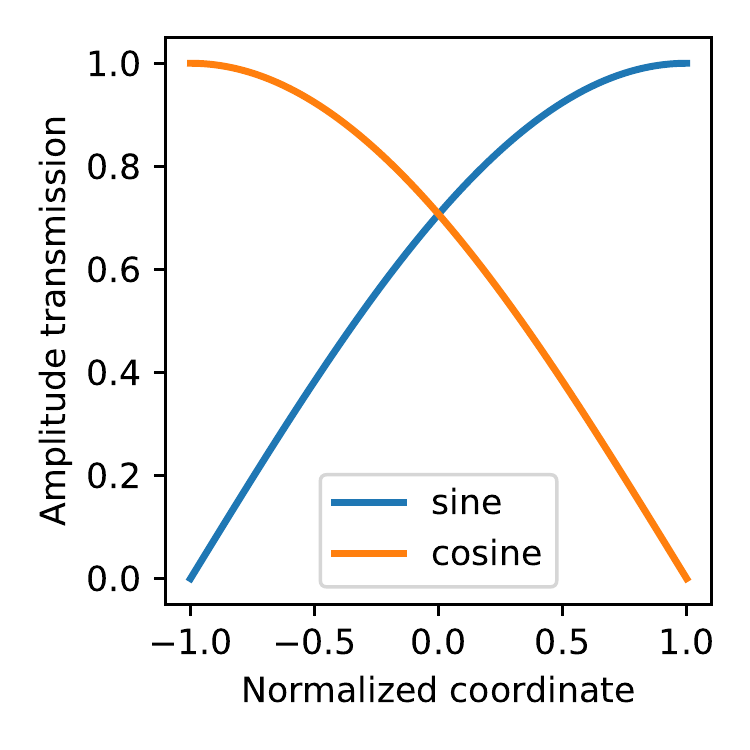}
	\end{center}
	\caption{The amplitude transmission of the two focal plane filters. The sine and cosine filters are mirrored versions of each other. }
	\label{fig:filters}
\end{figure*}

Mathematically the ODWFS can be described as a Fourier filter,
\begin{equation}
    E_1 =\mathcal{F}\left[ t \mathcal{F}\left[E_0\right] \right].
\end{equation}
Here $E_0$ is the input electric field, $\mathcal{F}$ is the Fourier transform operator, $t$ is the focal plane transmission mask and $E_1$ is the output electric field. The ODWFS uses two tranmission masks, one that measures wavefront variations in the x direction and one that measures variations in the y direction. In the definition of the filter masks, $k_x$ is used as focal plane coordinate which is normalized by the extent of the filter $k_m$. The normalized coordinates run from -1 to 1. For a single direction, we choose to describe the x-direction, the transmission mask creates two effective amplitude filters,
\begin{equation}
    t_+ = \sin{\left[ \frac{\pi}{4}\left(1+k / k_m\right) \right]},
\end{equation}
and,
\begin{equation}
    t_- = \cos{\left[ \frac{\pi}{4}\left(1+k / k_m\right) \right]}.
\end{equation}
Outside this range the filter has a transmission of zero. The filter transmission curves can be seen in Figure \ref{fig:filters}. We create the two perfectly mirrored focal plane transmission masks with the sine and cosine, which come from the projection onto the two polarization states. The normalized difference between the corresponding pupils can then be used to measure the wavefront.  The Fourier transform of both the sine and cosine filter is,
\begin{equation}
   \mathcal{F}\left[ \sin{\left[ \frac{\pi}{4}\left(1+k/k_m\right) \right]} \right] = \sqrt{8\pi}e^{i\pi/4}\left[\delta\left(x-\frac{\pi}{4k_m}\right) - i\delta\left(x +\frac{\pi}{4k_m}\right)\right],
\end{equation}
and
\begin{equation}
   \mathcal{F}\left[ \cos{\left[ \frac{\pi}{4}\left(1+k/k_m\right) \right]} \right] = \sqrt{8\pi}e^{-i\pi/4}\left[\delta\left(x-\frac{\pi}{4k_m}\right) + i\delta\left(x + \frac{\pi}{4k_m}\right)\right],
\end{equation}
respectively. Here $\delta$ is the Dirac $\delta$-function. The corresponding output electric fields are
\begin{equation}
    E_\pm = E_0 \ast \mathcal{F}\left[ t_\pm \right] = \sqrt{8\pi}e^{-\pm i\pi/4}\left[ E_0\left(x+x_0\right)-\pm i E_0\left(x-x_0\right) \right].
\end{equation}
For each filter $t_\pm$ there is an output electric field $E_\pm$. We also used $x_0 = \pi/4k_m$ to simplify the equations. The corresponding pupil intensities are,
\begin{equation}
    I_\pm = 8\pi\left[ I_0(x+x_0) + I_0(x-x_0) \pm 2\mathfrak{Re}\left(iE_0(x+x_0) E_0(x-x_0)^\dagger\right) \right].
\end{equation}
Here $I_\pm$ is the pupil intensity of the corresponding transmission mask $t_\pm$. The input electric field can be decomposed into its amplitude and phase, $E_0 = Ae^{i\phi}$, and again for clarification we use $f^{\pm}=f(x \pm x_0)$ to described a $\pm x_0$ shifted version of the variable $f$. Substituting these two relations gives us,
\begin{equation}
    \frac{I_\pm}{8\pi} = I_0^{+} + I_0^{-} \pm 2A^{+}A^{-}\mathfrak{Re}\left(ie^{i\phi^+-i\phi^-}\right)
    =I_0^{+} + I_0^{-} \pm 2A^{+}A^{-}\sin\left(\phi^+-\phi^-\right).
\end{equation}
From this we can see that the response of the ODWFS with sinusoidal filters is similar to the response of a lateral shearing interferometer with a very small shear. This can be simply understood because the sine/cosine filters are just very low-frequency amplitude gratings. The two pupil images can be used to estimate both the phase and amplitude in an identical way as in \cite{haffert2016generalised}. The normalized difference is used to estimate the phase, which for the sinusoidal filters is,
\begin{equation}
    \arcsin{\left[\frac{I_+ - I_-}{I_+ + I_-}\right]} \approx \left(\phi^+ - \phi^-\right).
\end{equation}
Here we assumed that the effects of amplitude variations and the pupil edges are negligible. The shifted phase difference can be related to its derivative by making use of a Taylor expansion,
\begin{equation}
   \phi^+ - \phi^- = 2x_0\frac{\partial\phi}{\partial x} + \mathcal{O}\left(\frac{\partial^3\phi}{\partial x^3} x_0^3\right).
\end{equation}
From this we can see that if the shift $x_0$ is small enough ($x_0 \ll 1$), the ODWFS with sinusoidal filters will measure the wavefront gradient. This is the case for all focal plane filters that are larger than 1 $\lambda/D_{tel}$. The optical gain, $2x_0$, is,
\begin{equation}
   2x_0 = \frac{\pi}{2 k_m} = \frac{\pi \lambda D_{tel}}{4\pi N \lambda_0}=\frac{ \lambda D_{tel}}{4 N \lambda_0}.
\end{equation}
Here we used $k_m=\frac{2\pi}{\lambda} N \lambda_0 / D_{tel}$ for the maximal spatial frequency with $N$ the number of Airy rings the focal plane filter encompasses. The size of the filter is defined at a central wavelength $\lambda_0$. Substituting the optical gain and the relation between phase and wavefront ($\phi=\frac{2\pi}{\lambda}W$), we get,
\begin{equation}
    \arcsin{\left[\frac{I_+ - I_-}{I_+ + I_-}\right]} \approx \frac{\pi}{2} \frac{D_{tel}}{N\lambda_0} \frac{\partial W}{\partial x}.
\end{equation}

\subsection{Extending the ODWFS to extended sources}
The previous section described the response of the ODWFS for a point source. The situation for an extended source is slightly more complicated. If we assume the extended source is incoherent, the pupil intensities will add,
\begin{equation}
    I_\pm = \int_S  I_0^{+} + I_0^{-} \pm 2A^{+}A^{-}\sin\left(\phi^+-\phi^-\right) dS.
\end{equation}
Here the integral is over the full source $S$. This can be split into two parts, the pupil illumination and the wavefront information,
\begin{equation}
    I_\pm = \langle I_0^{+} + I_0^{-} \rangle_S \pm \langle 2A^{+}A^{-}\sin\left(\phi^+-\phi^-\right)\rangle_S.
\end{equation}
If each of the incoherent source creates the same pupil illumination, we can take the amplitudes out of the average of the right term. And if the source ,including broadening due to turbulence, is small compared to the field of view of the filter, the normalised difference becomes,
\begin{equation}
   \frac{I_+ - I_-}{I_+ + I_-} \approx \langle \phi^+-\phi^-\rangle_S.
\end{equation}
From this we can see that the ODWFS measures the source intensity weighted wavefront.

\section{End-to-end simulations}
\label{sec:e2esims}  
In this section we will use end-to-end simulations to evaluate the performance of the ODWFS for LGS wavefront sensing. All simulations in this manuscript make use of the High Contrast Imaging for Python (HCIPy) library \cite{por2018high}. The main parameters of the simulation can be found in Table \ref{tab:simulation_parameters}.

\begin{table}[ht]
\caption{The simulations parameters.} 
\label{tab:simulation_parameters}
\begin{center}       
\begin{tabular}{|l|l|l|} 
\hline
\rule[-1ex]{0pt}{3.5ex}  \textbf{Parameter} & \textbf{Value} & \textbf{Comment} \\
\hline
\rule[-1ex]{0pt}{3.5ex}  $D_{tel}$ & 8.0 &   \\
\hline
\rule[-1ex]{0pt}{3.5ex}  wavelength & 589 nm & Sodium LGS  \\
\hline
\rule[-1ex]{0pt}{3.5ex}  \textbf{DM Parameters} &  & \\
\hline
\rule[-1ex]{0pt}{3.5ex}  Modes basis & Fourier modes & \\
\hline
\rule[-1ex]{0pt}{3.5ex}  Maximum frequency & 30 cycles / pupil & 899 modes in total with piston excluded  \\
\hline
\rule[-1ex]{0pt}{3.5ex}  \textbf{ODWFS Parameters} &  & \\
\hline
\rule[-1ex]{0pt}{3.5ex}  Number of sub-apertures & 32 pixels & \\
\hline
\rule[-1ex]{0pt}{3.5ex} Filter field of view & 4.5" & 150 $\lambda/D$ \\
\hline
\rule[-1ex]{0pt}{3.5ex}  \textbf{SHWFS Parameters} &  & \\
\hline
\rule[-1ex]{0pt}{3.5ex}  Number of sub-apertures & 41 lenslets & \\
\hline
\rule[-1ex]{0pt}{3.5ex}  Plate scale & 0.67" per pixel & \\
\hline
\rule[-1ex]{0pt}{3.5ex}  Lenslet field of view & 4" & \\
\hline

\end{tabular}
\end{center}
\end{table}

\subsection{Calibration of the wavefront sensors}
Both the ODWFS and the SHWFS are calibrated with a point source. This means that any error that is introduced because of the extended LGS will fold back into reconstruction errors of the wavefront. To calibrate the response of wavefront sensors, we used almost 4000 random patterns on the DM with a surface RMS of 200 nm. This data set was then divided into a training part (3000 samples) and a validation part (1000 samples). From the training data we derived the interaction matrix as,
\begin{equation}
    A = V_TS_T^T(S_TS_T^T + \alpha I)^{-1}.
\end{equation}
With $A$ the reconstruction matrix, $V_T$ the applied commands of the training data, $S_T$ the measured slopes of the training data, and $\alpha$ the regularization parameter. The optimal regularization parameter is chosen as the $\alpha$ which minimizes the residuals of the reconstructed coefficients in the validation dataset,
\begin{equation}
    \argmin_{\alpha} |V_V - A\left(\alpha\right) S_V|^2.
\end{equation}
Here, $V_V$ and $S_V$ are the applied commands and measured slopes of the validation data set.

\subsection{Using the ODWFS for LGS wavefront sensing}
Simulating a full LGS system is a complicated task both the up and downlink have to be simulated \cite{holzlohner2008physical}. During the uplink the LGS accumulates wavefront errors which create an aberrated illumination in the sodium layer. Because each part of the laser Point Spread Function (PSF) can excite a sodium atom in the sodium layer, each simulated pixel of the PSF should be regarded as an independent incoherent point source. Therefore, the laser PSF in the sodium layer can be seen as a probability density function of the emitting volume. Then during the downlink the LGS accumulates the wavefront errors of the atmosphere that are common with the science target.

\begin{figure*}
	\begin{center}
        \includegraphics[width = 0.5\textwidth]{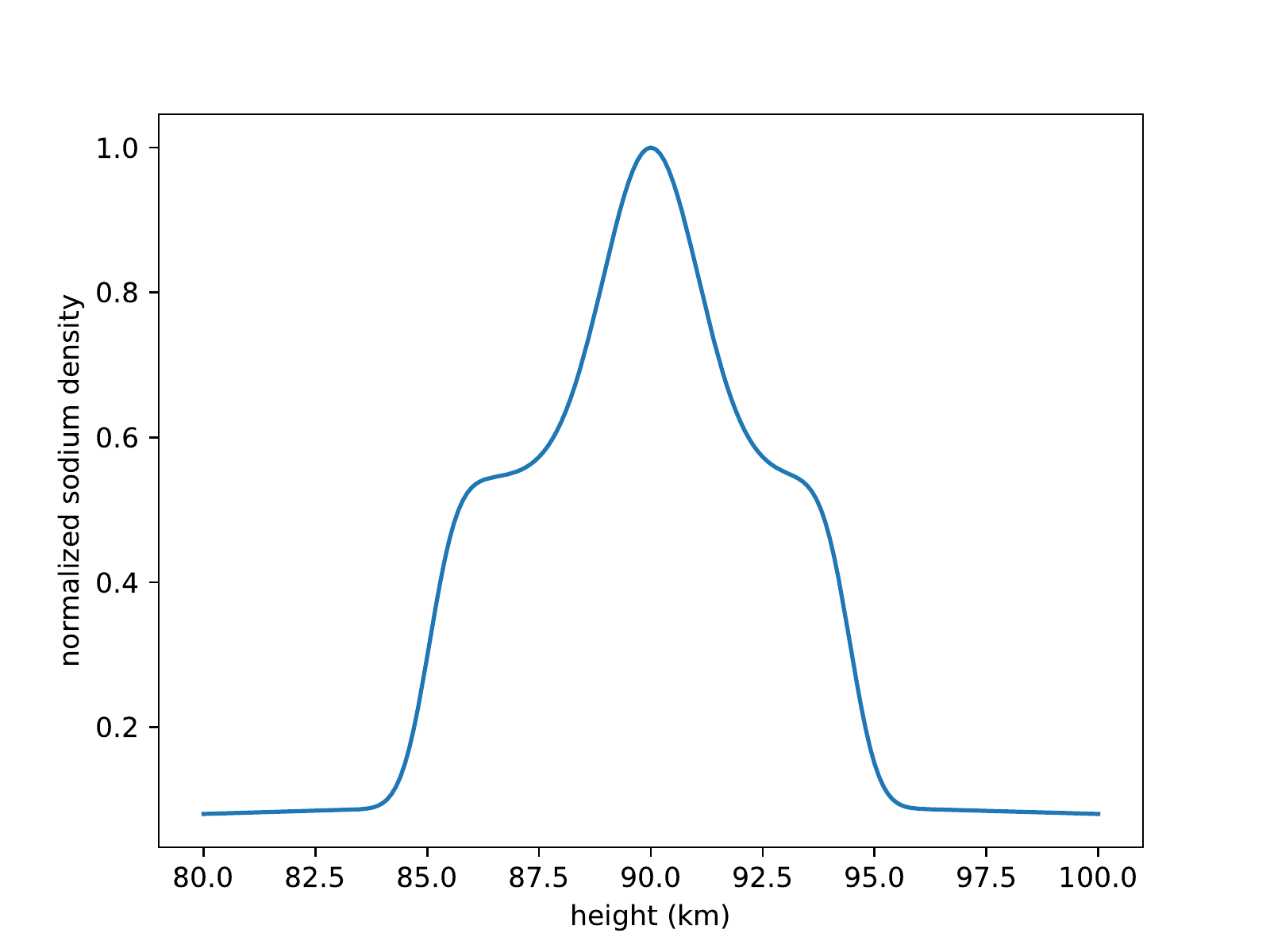}
	\end{center}
	\caption{The sodium density profile that is used for the simulations. There is a weak broad component that spans from 80 km to 100 km with a stronger broad component from 85 km to 95 km. And finally there is a Gaussian bump at 90 km with a sigma of 1.5 km.}
	\label{fig:sodium_distribution}
\end{figure*}

\begin{figure*}
	\begin{center}
       \includegraphics[width = \textwidth]{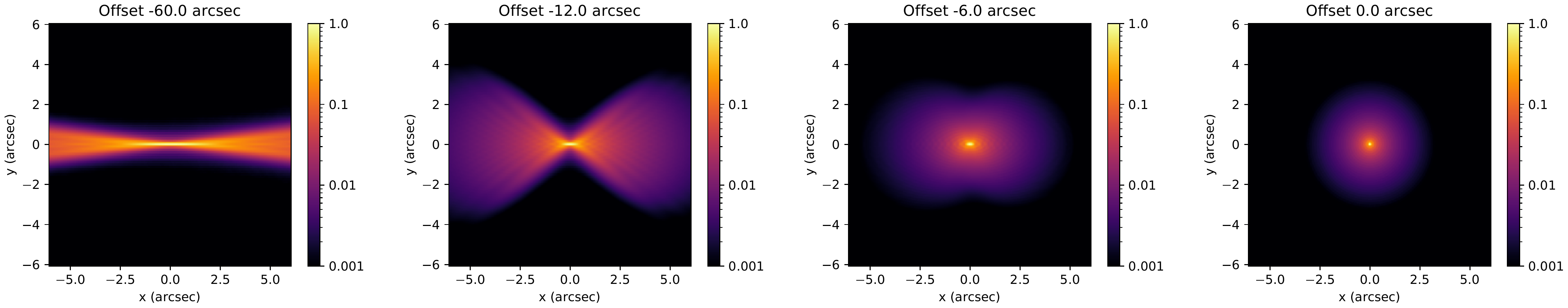}
	\end{center}
	\caption{These focal plane images on log-scale show the effect of the viewing angle on the structure of the LGS if viewed by a 8 meter class telescope. The image plane is conjugated to the mean height of the sodium layer. The LGS is more elongated as the viewing angle is increased. The LGS creates a butterfly pattern because each slice of the sodium layer will have an offset and a height difference that creates a focus offset. This butterfly effect scales with the diameter of the aperture.}
	\label{fig:lgs_elongation}
\end{figure*}

For this manuscript we are interested in the performance of the ODWFS for wavefront sensing of an extended object. The LGS lateral extent caused by the PSF in the sodium layer is much smaller than the elongation. Therefore, we decided to neglect the lateral extent and simulate the LGS as a line source. We have also ignored the propagation through a multi-layer atmosphere \cite{holzlohner2008physical} and assumed that all wavefront errors occur in the telescope pupil. To simulate the line source we split the sodium layer up into 21 independent layers. Each layer is assigned a power according to the density profile that can be seen in Figure \ref{fig:sodium_distribution}. We assumed that the LGS has an offset of 5 m with respect to the center of the pupil and the laser is shot perpendicular to the orientation of the pupil. The electric field of the each point source in the entrance pupil can be calculated from the impulse response function of free space propagation,
\begin{equation}
    E = \frac{e^{i\frac{2\pi}{\lambda}r}}{r}
\end{equation}
Here $r$ is the distance from the atoms that are excited by the laser to a position within the telescope pupil. Because $r$ depends on the position of the point source and the telescope pupil coordinate we naturally get the correct tilt and defocus error. The focal plane image of the LGS based on the sodium profile of Figure \ref{fig:sodium_distribution} can be seen in Figure \ref{fig:lgs_elongation}. The images were created with different viewing angles.

\begin{figure*}
	\begin{center}
        \includegraphics[width = \textwidth]{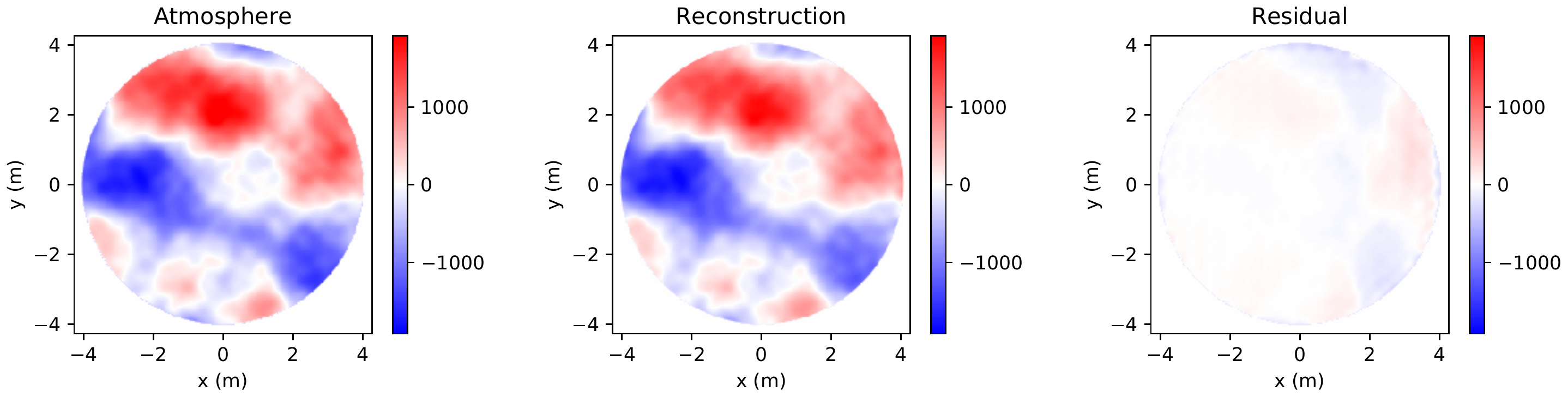}
	\end{center}
	\caption{An example of a reconstructed wavefront from the ODWFS with a LGS as input. The figure on the left shows the input wavefront and the figure in the middle shows the reconstructed wavefront. The difference between the input and reconstructed wavefront can be seen on the right. The rms of the residual is roughly 15 nm. The colorbar shows the scale of the wavefront in nm. }
	\label{fig:odwfs_reconstruction}
\end{figure*}

After the electric field of the LGS is created we pass it through a phase screen to create the atmospheric wavefront errors,
\begin{equation}
    E_0 = \frac{e^{i\frac{2\pi}{\lambda}r}}{r} e^{i\phi}.
\end{equation}
Each wavefront is then propagated through the wavefront sensor. We use a fully physical optics based simulation, so there are no geometric approximations used in the wavefront sensor optics. An example of a reconstructed wavefront from the ODWFS with an LGS can be seen in Figure \ref{fig:odwfs_reconstruction}. This shows that the ODWFS can create high-quality wavefront measurements.

\subsection{Comparison of the ODWFS and the SHWFS}
\begin{figure*}
	\begin{center}
        \includegraphics[width = \textwidth]{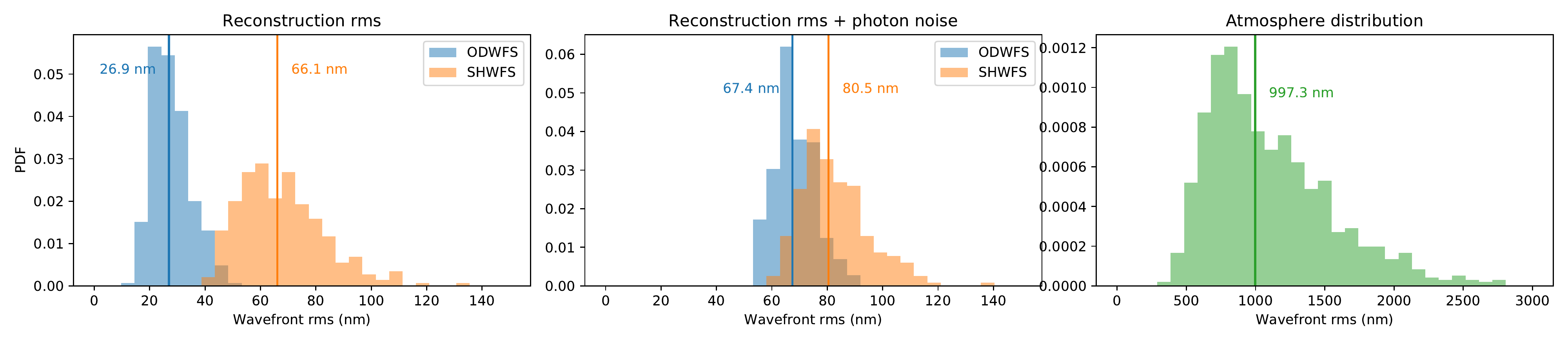}
	\end{center}
	\caption{The probability density function (PDF) as function of the residual wavefront rms. The left figure shows the rms of both the ODWFS (blue) and the SHWFS (orange) without photon noise, while the middle figure shows the PDF of the results with photon noise. The figure on the right shows the PDF of the rms of the incoming wavefronts. In all figures the median of the distribution is drawn with the solid vertical line together with the numerical value of the median. The ODWFS is performing better in both simulations.}
	\label{fig:compare_odwfs_shwfs}
\end{figure*}

The performance of the ODWFS was compared to that of the SHWFS by reconstructing 300 independent phase screen screens. The phase screens are generated from a Kolmogorov power spectrum with an $r_0$ of 15 cm and an $L_0$ of 25 m. To exclude the effects of spatial fitting errors, we project the phase screens onto the Fourier modes (Table \ref{tab:simulation_parameters}) before propagating the phase through the system. The histogram of both the atmosphere rms and the reconstructed rms are shown in Figure \ref{fig:compare_odwfs_shwfs}. The median reconstruction rms of the ODWFS is 27 nm compared to 66 nm for the SHWFS for a median input rms of 1 $\mu$m. The first histogram shows the reconstructed rms that contains the errors from the gain variations due to the spot elongation, substructure in the sodium density, non-linearities due to the large input phase and aliasing errors. The results of the Monte Carlo trials show that the ODWFS is more robust against these errors. The robustness against aliasing errors was already anticipated because the pyramid wavefront sensor, which is optically very similar, also has small aliasing errors \cite{verinaud2004nature}.

The photon noise sensitivity was also estimated by repeating the same numerical experiment but with the addition of photon noise. The results are shown in the middle histogram of Figure \ref{fig:compare_odwfs_shwfs}. The ODWFS still has a smaller total reconstruction error. The contribution of the photon noise can be estimated from, $\sigma_{p} = \sqrt{ \sigma^2-\sigma_{r}^2 }$. Here $\sigma_p$ is the rms of the photon noise, $\sigma$ the total noise and $\sigma_r$ the reconstruction noise. From this equation we estimate the photon noise for the ODWFS as $61$ nm and that of the SHWFS as $45$ nm. The ODWFS has a slightly worse photon noise sensitivity. The effects of read noise are not included, but these should be significantly smaller for the ODWFS than for the SHWFS because the ODWFS only uses 4 pixels per sub-aperture while the SHWFS uses 36 (6x6), and for the E-ELT this will increase even more to 100 (10x10) \cite{downing2014lgsd}.

\begin{figure*}
	\begin{center}
        \includegraphics[width = \textwidth]{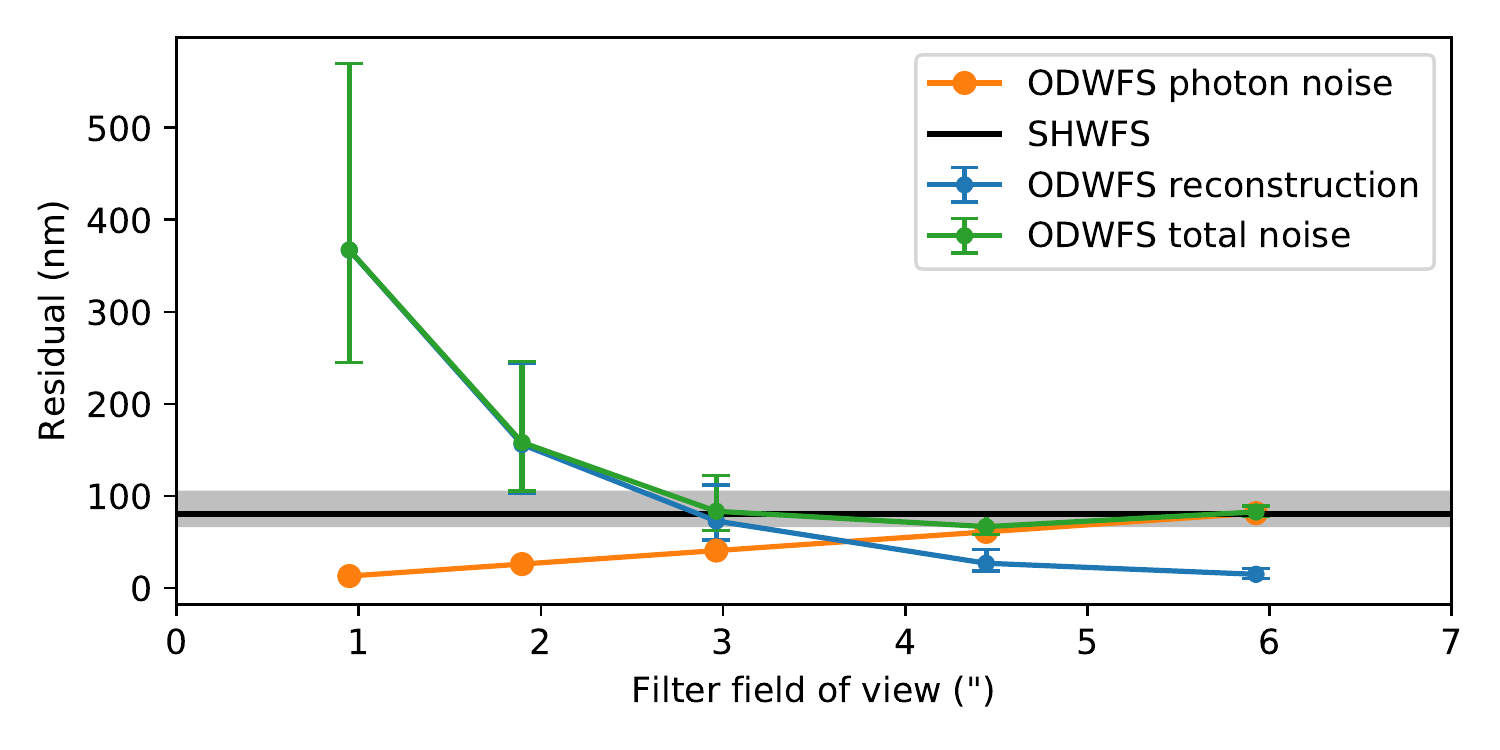}
	\end{center}
	\caption{The residual wavefront error as function of filter field of view. The orange line shows the photon noise and the blue line shows the reconstruction error. The green line shows the total error of the ODWFS. Each errorbar represents the 90\,\% confidence interval derived from 300 Monte Carlo trials. The black line is the median reconstruction quality of the SHWFS with photon noise (same data as the middle panel of Figure \ref{fig:compare_odwfs_shwfs}). The gray shaded area is the 90\,\% confidence interval of the SHWFS.}
	\label{fig:odwfs_truncation}
\end{figure*}

The photon noise sensitivity of the ODWFS depends on the slope of the focal plane filter \cite{haffert2016generalised}. A smaller filter will decrease the field of view, but also decrease the effects of photon noise. While the photon noise will decrease, the smaller field of view will truncate the LGS which will increase the reconstruction errors. We repeated the Monte Carlo trials for several filter sizes of the ODWFS to estimate the effects of the truncation. The results are shown in Figure \ref{fig:odwfs_truncation}. The reconstruction quality with a filter size of 3" to 6" is quite similar to that of the SHWFS, there is only a trade-off between the reconstruction quality and the photon noise. For smaller field-of-views the reconstruction error dominates in the ODWFS's error budget. The reconstruction error depends of course on the size of the LGS. In these simulations the LGS was shot from outside the pupil, if the LGS is shot from the M2 instead the effective size of the LGS will become smaller and a smaller field-of-view can be used. This could increase the performance of the ODWFS.

\section{Discussion and conclusion}
\label{sec:conclusion}
We have derived the response of a polarization based optical-differentiation wavefront sensor and showed that without any approximation on the input wavefront, we recover the wavefront derivative. For extended sources the response is more complicated and requires numerical simulations if the object is larger than the field of view of the wavefront sensor. If the extended object is smaller than the field of view, the sensor measures the source intensity averaged wavefront.

Numerical simulations show that the ODWFS performs equal or better than the SHWFS with an LGS. This shows there are no downsides to switch to an ODWFS for LGS wavefront sensing. The main advantage of the ODWFS is that it requires significantly less pixels, which enables E-ELT scale LGS wavefront sensing with current state-of-the-art EMCCD's such as the OCAM2K from First Light.

We have investigate the effects of elongation, truncation and photon noise on the performance of the ODWFS. In the future, we will similarly address issues for dynamic range and linearity when using the ODWFS with an LGS for open loop wavefront sensing. Additionally, we will investigate the ODWFS with a more realistic atmospheric model that better represents real life situations.

We expect to test the ODWFS with extended sources in a lab setup in 2021 as preparation for SIGHT (Sharpening Images using Guide stars at the Hale Telescope), which is an innovative, extremely compact laser guide star (LGS) adaptive optics (AO) pathfinder for which the ODWFS will be the main wavefront sensor.

\acknowledgments 
Support for this work was provided by NASA through the NASA Hubble
Fellowship grant \#HST-HF2-51436.001-A awarded by the Space Telescope
Science Institute, which is operated by the Association of Universities for Research in Astronomy, Incorporated, under NASA contract NAS5-26555.

\bibliography{report} 
\bibliographystyle{spiebib} 

\end{document}